\documentclass[aps,pre,twocolumn,floatfix,showpacs,showkeys]{revtex4}
\usepackage{amsmath,amsfonts,amssymb,mathrsfs,graphicx}
\begin{document}
%\tighten
%\draft
%@twocolumnfalse\endcsname
\title{A New Nonextensive Entropy}

\author{Fariel Shafee}
\affiliation{ Department of Physics\\ Princeton University\\
Princeton, NJ 08540\\ USA.} \email{fshafee@princeton.edu }

\begin {abstract}
We propose a new way of defining entropy of a system, which gives a
general form which may be nonextensive as Tsallis entropy, but is
linearly dependent on component entropies, like Renyi entropy, which
is extensive. This entropy has a conceptually novel but simple
origin and is mathematically easy to define by a very simple
expression, though the probability distribution resulting from
optimizing it gives rather complex, which is compared numerically
with the other entropies. It may, therefore, appear as the right
candidate in a physical situation where the probability distribution
does not suit any of the previously defined forms.

\end{abstract}

 \vspace*{1cm}
%]

\maketitle

\section{Introduction}

Entropy is a measure of disorder or randomness, and assumes its
maximal value when a system can be in a number of states randomly
with equal probability, and is minimally zero when the system is in
a given state, with no uncertainty in its description. Apart from
this common feature shared by all definitions of entropy at two ends
of the scale, variations are possible in particularizing the
functional form in between \cite{LA1}. They lead to different forms
of the probability distributions for states with different energies
or some other conserved attribute. Some turn up as extensive, where
the entropy of a combination of systems is simply the sum of the
entropies of the systems, as in the classical case of the Shannon
form, while others can be defined to be not so. Renyi entropy
\cite{RE1} is different from Shannon, yet extensive, and hence the
Shannon form is not unique with respect to the property of
extensivity.

Tsallis entropy \cite{TS1, TS2. PA1} has attracted a lot of
attention in recent years, not only on account of its conceptual and
theoretical novelty, but also because it can be shown in specific
physical cases \cite{BE1, CO1, WO1} to be the relevant form where
nonextensivity is expected on account of the interaction of the
combined subsystems. In the proper limiting case it reduces to the
standard Shannon entropy, indicating the consistency of the concept.

In this paper, however, we shall introduce entropy from a new
perspective, which too will bear semblance to the normal form in the
limit. We shall first present the rationale for this new definition
and compare it briefly with the forms already being used. Then we
shall find the form of the probability distribution for this
entropy, which we shall henceforth call s-entropy, as it will be
seen to be related to the concept of rescaling of the phase space.

\section{DEFINING THE NEW ENTROPY}
Let us consider a register of only one letter. Let ${p_i}$ be the
set of probabilities for each of the $N$ letters $A_i$ that can
occupy this position. We are here using the language of information
theory, as used for example, in the Shannon Coding Theorem, though
it is trivially extensible to states $i$ of a single state of an
ensemble where the individual systems can be in any $N$ states with
probabilities $p_i$.

Let us now consider a small deformation of the register to a new
size so that it can accommodate $q = 1+ \Delta q$ letters. The
probability that the whole new phase space is occupied by the letter
$p_i$ is now $p_i^q$ by the corresponding AND operation and hence
the probability that the new deformed cell is occupied by any of the
pure letters $A_i$ is

\begin{equation} \label{eq1}
N(q) =  \sum_i p_i^q
\end {equation}

For $q >1 $  this would give a shortfall from the original total
probability of unity for $q=1$. It is obvious that the shortfall,
which we denote by

\begin{equation} \label{eq2}
M(q) = 1- \sum_i p_i^q
\end {equation}

represents the total probability that the mixed cell has a mixture
of $A_i$ and some other $A_j$ fractionally, since the total
probability that the cell is occupied by one or more (fractional
included) letters must be unity. Hence the mixing probability $M(q)$
is actually a measure of the disorder introduced by increasing the
cell scale from unity to $1+ \Delta q$.

The introduction of fractional values of cell numbers can be taken
in the same spirit as defining the fractal (Hausdorff) dimensions of
curves, and in complex systems there have been studies of diffusion
\cite{BU1} and percolation in complex systems with effectively
fractional dimensions for fluids where the special geometric
constraints translate into a change in the dimension of the
corresponding space to an apparently nonintuitive fractional
dimension. In coding theories for optimal transmission of
information \cite{NI1}, we come across Huffmann coding, where the
optimum alphabet size may be formally a fraction, though for
practical purposes it may be changed to the nearest higher integer.
In probabilistic optimization, we may therefore consider a
fractional size of the registrar, or equivalently, an integral
number of cells in the registrar with fractional sized cells to
accommodate a given amount of information. Probabilistic
optimization in place of the deterministic parameterization of
classical Shannon information theory \cite{NI1} becomes inevitable
in quantum computing contexts, and hence our use of the fractional
cell sizes may be a classical precursor of the inevitable departure
from stringent Shannon-type concepts.

For an alphabet of $m$letters we define the entropy from the
information content of the registrar by

\begin{equation} \label{eq3}
m^{S(q) \Delta q}  = m^{(M(q+\Delta q)- M(q))}
\end {equation}

so that the entropy indicates an effective change in the mixing
probability due to an infinitesimal change in the cell-size of the
registrar.

This leads to

\begin{equation} \label{eq4}
S(q)  = dM(q)/dq
\end {equation}

In other words

\begin{equation} \label{eq5}
S(q) = - \sum_i p_i^q \log p_i
\end {equation}

{\bf We have some material at the other place.} This differential
form is analogous to but different from the Tsallis form

\begin{equation} \label{eq6}
S_T(q) = - \sum_i (1- p_i^q)/ (1-q)
\end {equation}

where there is an apparent singularity at $q=1$ which is the Shannon
limit. The difference between the Tsallis expression and ours
becomes clearer if we express entropy as the expectation value of
the (generalized or ordinary) logarithm.

\begin{equation} \label{eq7}
S_T(q) = <Log_q p>
\end {equation}

where the generalized q-logarithm is defined as

\begin{equation} \label{eq8}
Log_q  p_i  = 1- p_i^{(q-1)}/ ( 1- q)
\end {equation}

The expectation value is defined in terms of the simple probability
distribution
\begin{equation} \label{eq9}
<O> =  \sum_i p_i O_i
\end {equation}

In our case we define the expectation value with respect to the
deformed probability corresponding to the extended cell, while
keeping the usual logarithm

\begin{equation} \label{eq10}
S_s(q) = < \log p >_q
\end{equation}

with

\begin{equation} \label{eq11}
<O>_q = \sum_i p_i^q O_i
\end{equation}

In the limit $q \rightarrow 1$  $Log_q$ approaches the normal
logarithm, and hence Tsallis entropy coincides with Shannon entropy
and also as $p_i^q \rightarrow  p_i$ we too get the normal Shannon
entropy.

The Renyi entropy is defined by

\begin{equation} \label{eq12}
S_R(q) = \log ( \sum_i p_i^q)/(1-q)
\end{equation}

Like Shannon entropy this one is also extensive, i.e. simply
additive for two subsystems for any value of $q$. To get Shannon
entropy uniquely one needs \cite{KI1} a slightly different
formulation of the extensivity axiom

\begin{equation}\label{eq13}
S_{1+2} = S_1 + \sum_i p_{1i}  S_2(i),
\end{equation}

where $S_2(i)$ is the entropy of subsystem $2$ given subsystem $1$
is in state $i$.

\section{PROBABILITY DISTRIBUTION FOR THE NEW ENTROPY}

The $p_i$ can be obtained in terms of the energy of the states, or
possibly also other criteria in the usual way by maximizing the
entropy with constraints

\begin{equation}\label{eq14}
\sum_i p_i -1=0
\end{equation}

and

\begin{equation}\label{eq15}
\sum_i p_i E_i - U = 0
\end{equation}.

The solution of the optimization equation gives for energy $E_i$ the
probability $p_i$

\begin{equation}\label{eq16}
p_i = (\frac{ - q W(z)}{(a+ b E)(q-1)})^{1/(1-q)}
\end{equation}

where
\begin{equation}\label{eq17}
z=-e^{(q-1)/q} (a+ b E) (q-1)/q
\end{equation}

and W(z) is the Lambert function defined by

\begin{equation}\label{eq18}
z= w e^w
\end{equation}

Here $a$ and $b$ are constants coming from the Lagrange's
multipliers for the two constraints and are related to the overall
normalization and to the relative scale of energy, i.e. to
temperature ( $1/(kT)$) as in the Shannon case where we get the
Gibbs expression for $p_i$. In the Tsallis case $p_i$ has the
well-known value

\begin{equation}\label{eq19}
p_i = ( a + b (q-1)E)^{1/(1-q)}
\end{equation}

which is easily seen to reduce to Shannon form for $q \rightarrow
1$.

After some algebra it can be shown that this form reduces to the
Shannon form for $q \rightarrow 1$.

The nonextensivity of Tsallis entropy is seen easily by expanding

\begin{eqnarray}\label{eq2021}
S^T{}_{1+2} = - \sum_{ij} p_i p_j  (1- p_i{}^q p_j{}^q)/(1-q)^2 \\
= S^T{}_1+ S^T{}_2 + (1-q) S^T{}_1 S^T{}_2
\end{eqnarray}

For Renyi entropy we have the simple additive relation

\begin{equation}\label{eq22}
S^R{}_{1+2}= S^R{}_1 + S^R{}_2
\end{equation}

In case of the new entropy

\begin{equation} \label{eq23}
S^s{}_{1+2} = S^s{}_1 + S^s{}_2 + M_2(q)S^s{}_1 + M_1(q)S^s{}_2
\end{equation}

where the $M_a$ are the mixing probability of states for subsystem
$a$ as defined in Eq.2.

\section{NUMERICAL COMPARISON}

In Fig.1 we show the variation of the probability function for
different $E$  at  different $q$ values.

\begin{figure}[ht!]
\includegraphics[width=8cm]{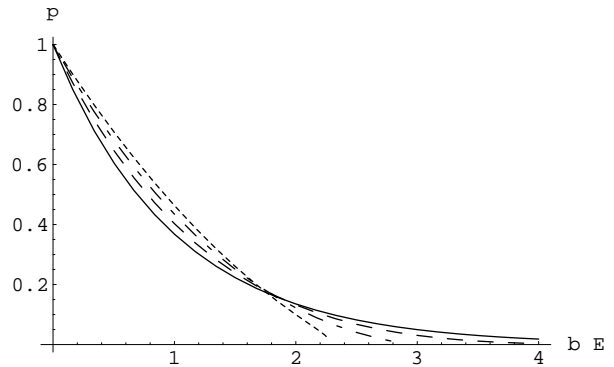}
\caption{\label{fig1}Comparison of the pdf for the new entropy for
values of $q=1,1.1,1.2$ and $1.3$. The solid line is for $q=1$, i.e.
the Gibbs exponential distribution and the lines are in the order of
$q$}
\end{figure}

We note that the pdf drops increasingly rapidly for higher values of
$q$, and is quite different in shape and in magnitude at high energy
values from the Gibbs exponential distribution. A variation of even
$10\%$ from the standard value of $q=1$ can cause a quite
discernible change in the pdf and should be observable in
experimental contexts fairly easily . At $q=1.3$, the shape is
almost linear.

In Fig.2 and Fig.3 we show the comparison of Tsallis pdf and the pdf
for the new entropy for the same values of $q$, 1.1 in the former
and 1.3 in the latter. We notice that for larger $q$ values the new
entropy gives much stiffer probability functions departing
substantially from the Tsallis pdf's.

\begin{figure}[ht!]
\includegraphics[width=8cm]{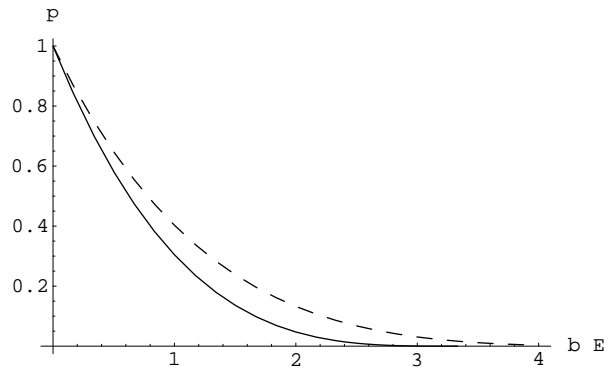}
\caption{\label{fig2}Comparison of pdf's for Tsallis nonextensive
entropy and the new entropy presented here, for $q=1.1$}
\end{figure}

\begin{figure}[ht!]
\includegraphics[width=8cm]{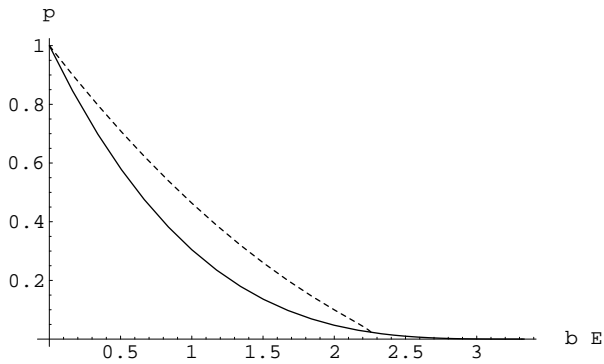}
\caption{\label{fig3}The same as Fig. 2, but for a higher $q=1.3$}
\end{figure}

\section{CONCLUSIONS}

We see that the new entropy presented here based on the simple
concept of the amount of mixing of states freedom introduced per
unit cell of phase space leads to a nonextensive form different from
any of the presently studied entropies. It leads to a complicated,
but still integrable form of the pdf which departs substantially
from Tsallis entropy. This entropy is also nonextensive in a fashion
different from Tsallis entropy, though like Tsallis it too becomes
extensive trivially in the limit $q \rightarrow 1$, as expected.

It would now be interesting to find a physical situation where such
an entropy arises from first principles, though like some initial
phenomenological studies of Tsallis entropy it can be also used as a
parametrization scheme with $q$ as a parameter to fit experimental
data. The stiffness of any data may point to its preferability to
Tsallis-type entropies.

\end{document}